\documentclass[preprint,showpacs,preprintnumbers,amsmath,amssymb]{revtex4}

\usepackage{graphicx}% Include figure files
\usepackage{dcolumn}% Align table columns on decimal point
\usepackage{bm}% bold math

\begin{document}

\preprint{}

\title{Quasilocal mass in general relativity}% Force line breaks with \\

\author{Mu-Tao Wang}
% \altaffiliation[Also at ]{Columbia University,}%Lines break automatically or can be forced with \\
 \affiliation{%
Columbia University, Department of Mathematics, 2990 Broadway, New York, New York 10027, USA\\
}%
\author{Shing-Tung Yau}%
\affiliation{%
Harvard University, Department of Mathematics, One Oxford Street, Cambridge, Massachusetts 02138, USA\\
}%

\date{October 8, 2008}% It is always \today, today,

\begin{abstract}
 There have been many attempts to define the notion of quasilocal mass for a spacelike 2-surface in spacetime by the Hamilton-Jacobi analysis. The essential difficulty in this approach is to identify the right choice of the background configuration to be subtracted from the physical
  Hamiltonian. Quasilocal mass should be nonnegative for surfaces in general spacetime and zero for surfaces in flat spacetime. In this letter, we propose
  a new definition of gauge-independent quasilocal mass and prove that it has the desired properties.
\end{abstract}

\maketitle

\section{\label{sec:level1} Introduction}

As is well known, by the equivalence principle there is no well-defined concept of energy density in general relativity. On the other hand, when there is asymptotic symmetry, concepts of total energy and momentum can be defined. This is called the ADM energy-momentum and the Bondi energy-momentum
when the system is viewed from spatial infinity and null infinity, respectively. These concepts are fundamental in general relativity
and have been proven to be natural and to satisfy the important positivity condition in the work of Schoen-Yau \cite{sy}, Witten \cite{wi}, etc.  However,
there are limitations to such definitions if the physical system is not isolated and cannot quite be viewed
from infinity where asymptotic symmetry exists. It was proposed more than 40 years ago to measure the energy of a system by enclosing
 it with a membrane, namely a closed spacelike 2-surface, and then attach to it an energy-momentum 4-vector. It is natural to expect that the 4-vector will depend only on the induced metric, the second fundamental form,
and the connection on the normal bundle of the surface embedded in spacetime. This is
the idea behind the definition of quasilocal mass of this surface. Obviously there are a few conditions the quasilocal mass has to satisfy:
Firstly, the ADM or Bondi mass should be recovered as spatial or null infinity is approached. Secondly, the correct
limits need be obtained when the surface converges to a point. Thirdly and most importantly, quasilocal mass must be nonnegative in
general and zero when the ambient spacetime of the surface is
the flat Minkowski spacetime. It should also behave well when the
spacetime is spherically symmetric. Many proposals were made by Hawking \cite{ha}, Penrose \cite{pe}, etc. The most promising
one was proposed by Brown-York \cite{by} where they motivated their definition by using the Hamiltonian formulation of general relativity (see also Hawking-Horowitz \cite{hh}). They found interesting local quantities from which the definition of quasilocal mass was extracted.
Their definition depends on the choice of gauge along the 3-dimensional spacelike slice which the surface bounds. It has the right asymptotic behavior but is
not positive in general. Shi-Tam \cite{st} proved that it is positive when the 3-dimensional slice is time symmetric. Motivated by geometric consideration, Liu-Yau \cite{ly} (see also Kijowski \cite{ki}, Booth-Mann\cite{bm}, and Epp \cite{ep}) introduced a mass which is gauge independent, and
proved that it is always positive. However, it was pointed out by \'{O}Murchadha \textit{et al.} \cite{ost} that the Liu-Yau mass
can be strictly positive even when the surface is in a flat spacetime.
In this letter, we explore more in the direction of the Hamilton-Jacobi analysis of Brown-York.  Combining some ideas from Liu-Yau, we define a quasilocal mass which is gauge independent and nonnegative. Moreover, it is zero whenever the surface is in the flat Minkowski spacetime. We believe that the present definition satisfies all the requirements necessary for a valid definition of quasilocal mass, and it is likely to be the unique definition that satisfies all the desired properties.

\section{Hamiltonian formulation revisited}
Consider a spacetime region $M$ that is foliated by a family of spacelike hypersurface $\Omega_t$ for $t$ in the time interval $[t', t'']$. The boundary of $M$ consists of $\Omega_{t'}$, $\Omega_{t''}$, and ${}^3B$.
Let $u^\mu$ denote the future pointing timelike unit normal to $\Omega_t$. Assume $u^\mu$ is tangent to ${}^3B$.
  Denote the boundary of $\Omega_t$ by $\Sigma_t$ which is the intersection of $\Omega_t$ and ${}^3B$. Let $v^\mu$ denote the outward pointing spacelike
  unit normal of $\Sigma_t$ such that $u_\mu v^\mu=0$. Denote by $k$ the trace of the 2-dimensional extrinsic curvature of $\Sigma_t$ in $\Omega_t$ in
  the direction of $v^\mu$.
Denote the Riemannian metric, the extrinsic curvature, and the trace of the extrinsic curvature on $\Omega_t$ by $g_{\mu\nu}$, $K_{\mu\nu}=\nabla_{\mu} u_\nu$,
and $K=g^{\mu\nu}K_{\mu\nu}$, respectively.
Let $t^\mu$ be a timelike vector field satisfying
$t^\mu\nabla_\mu t=1$.  $t^\mu$ can be decomposed into the lapse function and shift vector $t^\nu=Nu^\nu+N^\nu$.
Let $S$ denote the action for $M$, then the Hamiltonian at $t''$ is given by $H=-\frac{\partial S}{\partial t''}$.
The calculation in Brown-York \cite{by} (see also Hawking-Horowitz \cite{hh}) leads to
\begin{equation}\label{hamil1}H=-\frac{1}{8\pi}\int_{\Sigma_{t''}} [Nk-N^\mu v^\nu(K_{\mu\nu}-Kg_{\mu\nu})]\end{equation} on a solution $M$ of the Einstein equation.
To define quasilocal energy, we need to find a reference action $S_0$ that corresponds to fixing the metric on ${}^3 B$ and compute the corresponding reference Hamiltonian $H_0$. The energy is then $E=-\frac{\partial}{\partial t''}(S-S_0)= H-H_0.$
In Brown-York's prescription, the reference is taken to be an isometric embedding of $\Sigma$ into $\mathbb{R}^3$, considered as a flat 3-dimensional slice with $K_{\mu\nu}=0 $ in a flat spacetime.  Choosing $N=1$ and $N^\mu=0$, the Brown-York quasilocal energy is
$\frac{1}{8\pi}\int_{\Sigma} (k_0-k)$.  References such as surfaces in the light cones (\cite{bly} and \cite{la}) and other conditions \cite{ki} have been proposed. However, the Brown-York energy for the examples \cite{ost} of surfaces in the Minkowski are in general
 non-zero for all these references. We shall define quasilocal energy using general isometric embeddings into $\mathbb{R}^{3,1}$ as reference configurations.
 In particular, the $t^\nu$ is obtained by transplanting a Killing vector field in $\mathbb{R}^{3,1}$ to the physical spacetime through the embedding.

\section{\label{sec:level2} Definition of quasilocal energy in the canonical gauge}
Suppose $\Sigma$ is a spacelike surface in a time orientable spacetime, $u^\nu$ is a future-pointing timelike unit normal,
and $v^\nu$ is a spacelike unit normal with $u^\nu v_\nu=0$ along $\Sigma$. We assume $v^\nu$ is the outward normal of a spacelike hypersurface $\Omega$
 that is defined locally near $\Sigma$.  Consider the 4-vector field
\begin{equation}\label{emd} ku^\nu+ v^\mu(K^{\nu}_\mu-K\delta^{\nu}_\mu)\end{equation} along $\Sigma$.
The definition of this vector field only depends on the two normals $u^\nu$ and $v^\nu$ along $\Sigma$.
The normal component (with respect to $\Sigma$) of (\ref{emd}) is $j^\nu=k u^\nu-pv^\nu $, where $p=K-K_{\mu\nu} v^\mu v^\nu$. $j^\nu$,
 as well as the \textit{mean curvature vector field}, $h^\nu=-kv^\nu+pu^\nu $, are defined independent of the choice of gauge $u^\nu$ and $v^\nu$.

 Consider a reference isometric embedding $i:\Sigma \hookrightarrow \mathbb{R}^{3,1}$ of $\Sigma$. Fix a constant timelike
  unit vector $t_0^\nu$ in $\mathbb{R}^{3,1}$, and choose a preferred pair of normals ${u}_0^\nu$ and ${v}_0^\nu$ along $i(\Sigma)$ in the following way:
   Take a spacelike hypersurface ${\Omega}_0$ with $\partial {\Omega}_0=i(\Sigma)$ and such that the outward pointing spacelike unit normal ${v}_0^\nu$
   of $\partial \Omega_0$ satisfies $(t_0)_\nu {v}_0^\nu=0$.
 Let ${u}_0^\nu$  be the future pointing timelike unit normal of ${\Omega}_0$ along $i(\Sigma)$. We can similarly form $k_0 u_0^\nu+ v_0^\mu((K_0)^{\nu}_\mu-K_0\delta^{\nu}_\mu)$ in terms of the corresponding geometric quantities on $\Omega_0$ and $i(\Sigma)$.
  $({u_0}^\nu, {v_0}^\nu)$ along $i(\Sigma)$ in $\mathbb{R}^{3,1}$ is the reference normal gauge we shall fix, and it depends on the choice of the pair $(i, t_0^\nu)$.

  When the mean curvature vector $h^\nu$ of $\Sigma$ in $M$ is spacelike, a reference isometric embedding $i:\Sigma\hookrightarrow\mathbb{R}^{3,1}$ and $t_0^\nu\in \mathbb{R}^{3,1}$ determine a canonical future-pointing timelike normal vector field $\bar{u}^\nu$ in $M$ along $\Sigma$.
 Indeed, there is a unique $\bar{u}^\nu$ that satisfies
\begin{equation}\label{canon}h_\nu \bar{u}^\nu=(h_0)_\nu{u_0}^\nu\end{equation} where $h_0^\nu$ is the mean curvature vector of $i(\Sigma)$ in $\mathbb{R}^{3,1}$. Physically, (\ref{canon}) means the expansions of $\Sigma\subset{M} $ and $i(\Sigma)\subset \mathbb{R}^{3,1}$
along the respective directions $\bar{u}^\nu$ and ${u_0}^\nu$ are the same.
This condition corresponds to fixing the metric on ${}^3B$ up to the first order in choosing the reference Hamiltonian in II.
$\bar{u}^\nu$ shall be called the {\it canonical gauge} with respect to the pair $(i, t_0^\nu)$. Take $\bar{v}^\nu$ to be the spacelike normal
vector that is orthogonal to $\bar{u}^\nu$ and satisfies $\bar{v}^\nu h_\nu<0$, and take a spacelike hypersurface $\bar{\Omega}$ in $M$ such that
 $\bar{v}^\nu$ is the outward normal. We can similarly form $\bar{k}\bar{u}^\nu+ \bar{v}^\mu(\bar{K}^{\nu}_\mu-\bar{K}\delta^{\nu}_\mu)$,
  where $\bar{K}^{\nu}_\mu$, $\bar{K}$, and $\bar{k}$ are the corresponding data on $\bar{\Omega}$. The trace of the 2-dimensional extrinsic curvature $\bar{k}$ of $\Sigma$ with respect to $\bar{v}^\nu$ is then given by
$\bar{k}=-\bar{v}^\nu h_\nu>0$.

4-vectors in  $\mathbb{R}^{3,1}$  and $M$, along $i(\Sigma)$ and $\Sigma$ respectively, can be identified through
\begin{equation}\label{iden}{u_0}^\nu\rightarrow \bar{u}^\nu, {v_0}^\nu\rightarrow \bar{v}^\nu,\end{equation} and the identification of tangent vectors
on $i(\Sigma)$ and $\Sigma$.

  The {\it quasilocal energy of $\Sigma$ in the canonical gauge with respect to $(i, t_0^\nu)$}  is defined to be
\begin{equation}\label{qlm_can}\frac{1}{8\pi}\int_\Sigma [\bar{k}\bar{u}^\nu+ \bar{v}^\mu(\bar{K}^{\nu}_\mu-\bar{K}\delta^{\nu}_\mu)-k_0{u}_0^\nu-{v}_0^\mu((K_0)^{\nu}_\mu-K_0\delta^{\nu}_\mu)] ({t}_0)_\nu, \end{equation}
where $\bar{u}^\nu$ is determined by (\ref{canon}), the identification (\ref{iden}) is used, and $t_0^\nu$ in $\mathbb{R}^{3,1}$ is identified with $N_0\bar{u}^\nu+N_0^\nu$ in $M$. In terms of the lapse $N_0$ and shift $N_0^\nu$, the quasilocal energy is \begin{equation}\frac{1}{8\pi}\int_\Sigma (k_0-\bar{k})N_0-(v_0^\mu(K_0)_{\mu\nu}-\bar{v}^\mu\bar{K}_{\mu\nu})N_0^\nu.\end{equation}

The mean curvature vector $h^\nu$ being spacelike is equivalent
to $\rho\mu>0$ where $\rho$ and $\mu$ are the expansion along the future and past outer null-normals of $\Sigma$.
When the image of the reference embedding lies in a flat space slice in $\mathbb{R}^{3,1}$, we have $t_0^\nu=u_0^\nu$. On the other hand,
the canonical gauge is $\bar{u}^\nu=\frac{1}{\sqrt{8\rho\mu}} j^\nu$, and we derive that $\bar{k}=\sqrt{8\rho\mu}$.  In this case, (\ref{qlm_can})
recovers the Liu-Yau quasilocal mass $\frac{1}{8\pi}\int_{\Sigma} ({k_0}-\sqrt{8\rho\mu})$.
\section{Admissible pairs}

Unlike Brown-York or Liu-Yau, we do not require the surface $\Sigma$ to have positive Gauss (intrinsic) curvature and apply the embedding theorem of Weyl. Instead, we prove a uniqueness and existence theorem of isometric embeddings into the Minkowski space under a more general convexity condition.

\textit{Definition 1 - Let $t_0^\nu$ be a  constant timelike unit vector in $\mathbb{R}^{3,1}$. A closed surface $\Sigma$ in $ \mathbb{R}^{3,1}$ is said to have convex shadow in the direction of $t_0^\nu$
if the projection of $\Sigma$ onto the orthogonal complement $\mathbb{R}^3$ of $t_0^\nu$ is a convex surface.}

The set of isometric embeddings with convex shadows is  parametrized by functions satisfying a convexity condition.

\textit{Theorem 1 -Let $\sigma_{ab}$ be a Riemannian metric on a two-sphere $\Sigma$. Given any function $\tau$ on $\Sigma$ with
\begin{equation}\label{convex}\kappa+(1+\sigma^{ab} \nabla'_a \tau \nabla'_b \tau)^{-1}\frac{\det(\nabla'_a\nabla'_b \tau)}{\det \sigma_{ab}}>0\end{equation} where $\kappa$ is the Gauss curvature and $\nabla'$ is the covariant derivative of the metric $\sigma_{ab}$. Then there exists a unique spacelike embedding $i:\Sigma\hookrightarrow\mathbb{R}^{3,1}$ such that the time function restricts to $\tau$ on $i(\Sigma)$ and the induced metric on $i(\Sigma)$ is $\sigma_{ab}$.}

 Uniqueness is dealt with first. Suppose there are two such isometric embeddings $i_1$ and $i_2$ with the same time function $\tau$. It is not hard to check that the condition (\ref{convex}) implies the projections of $i_1(\Sigma)$ and $i_2(\Sigma)$ onto the orthogonal complement of the time direction are isometric as convex surfaces in $\mathbb{R}^3$. By Cohn-Vossen's rigidity theorem, the projections are congruent by a rigid motion of $\mathbb{R}^3$. Since they have the same time functions, $i_1(\Sigma)$ and $i_2(\Sigma)$ are congruent by a Lorentizian rigid motion of $\mathbb{R}^{3,1}$. To prove existence, condition (\ref{convex}) is shown to imply that the metric $\sigma_{ab}+\nabla'_a \tau \nabla'_b\tau$ has positive Gauss curvature and thus can be isometrically embedded into $\mathbb{R}^3$. We may assume this $\mathbb{R}^3$ is a space-slice in the Minkowski space, so the induced metric on the graph of $\tau$ in $\mathbb{R}^{3,1}$ is exactly $\sigma_{ab}$. This completes the proof of Theorem 1.

In order to recognize a surface in the Minkowski space, we solve a Dirichlet boundary value problem for Jang's equation. Given a hypersurface $(\Omega, g_{ij}, K_{ij})$ in $M$, Jang's equation seeks for a solution $f$ of
\begin{equation}\label{jang1}\sum_{i, j=1}^3 (g^{ij}-\frac{D^i f D^j f}{1+g^{ij} D_if D_j f})(\frac{D_i D_j f}{(1+g^{ij}D_i fD_j f)^{1/2}}-K_{ij})=0, \end{equation} where $D$ is the covariant derivative of $g_{ij}$.  The graph of $f$ in the space $\Omega\times \mathbb{R}$ is denoted by $\widetilde{\Omega}$. In this article, we are interested in the case when $\partial \Omega=\Sigma$ and the prescribed value of $f$ on the boundary $\Sigma$ is given. Notice that if $\Sigma$ is in $\mathbb{R}^{3,1}$ and if we take the time function $\tau$ as the boundary value to solve Jang's equation, then $\widetilde{\Omega}$ will be a flat domain in $\mathbb{R}^3$.

 \textit{Definition 2 - Consider a spacelike 2-surface  $\Sigma$ in a time orientable spacetime $M$, an isometric embedding $i:\Sigma \hookrightarrow \mathbb{R}^{3,1}$, and a constant timelike unit vector $t_0^\nu\in \mathbb{R}^{3,1}$. Let $\tau$ denote the time function restricted to $i(\Sigma)$.  $(i, t_0^\nu)$ is said to be an {\textit admissible pair} for $\Sigma$ if the following conditions are satisfied:
 \vskip 5pt
 \noindent(A) $i$ has convex shadow in the direction of $t_0^\nu$.
 \vskip 5pt
\noindent(B) $\Sigma$ bounds a spacelike domain $\Omega$ in $M$ such that
Jang's equation (\ref{jang1}) with the Dirichlet boundary data $\tau$ is solvable on $\Omega$ (with possible apparent horizons in the interior).
\vskip 5pt
\noindent(C) Suppose $f$ is the solution of Jang's equation in (B) and $v^\nu$ is the
outward unit normal of $\Sigma$ that is tangent to $\Omega$, and
$u^\nu$ is the future-pointing timelike normal of $\Omega$ in $M$. Consider the new gauge given by
$u'^\nu=\sinh\phi v^\nu+\cosh\phi u^\nu$ and $v'^\nu=\cosh\phi v^\nu+\sinh\phi u^\nu$,
where $\sinh\phi=\frac{f_v}{\sqrt{1+\sigma^{ab}\nabla'_a \tau\nabla'_b \tau}}$, and $f_v$ is the normal derivative of $f$ in the direction $v^\nu$. We require that
\begin{equation}\label{pos_int}k'N_0-N_0^\nu v'^\mu K'_{\mu\nu}>0,\end{equation} where $k'$, $g'_{\mu\nu}$, and $K'_{\mu\nu}$ are the corresponding data on the new three dimensional spacelike domain $\Omega'$ spanned by $v'^\nu$, and $N_0$ and $ N_0^\nu$ are the lapse function and shift vector of $t_0^\nu=N_0 u_0^\nu+N_0^\nu$.}

 Remark 1 - By a barrier argument, we show that $\Omega$ satisfies (B) if on $\Sigma$, $k>\frac{1}{t^at_a(1+t^at_a)}(K_{ab} t^at^b)+K_{ab}u^au^b$ where $u^a$ is a two-vector such that $t^a u_a=0$ and $u^a u_a =1$, and $t^a=\pi^a_\nu t_0^\nu$ is  the projection of $t_0^\nu$ onto $\Sigma$. Also by elliptic estimates, (C) will be satisfied if (\ref{pos_int}) holds for $u^\nu$ and $v^\nu$, and $\sigma^{ab} \nabla'_a \tau \nabla'_b \tau$ is small enough. In particular, if $\Sigma$ has positive Gauss curvature and spacelike mean curvature vector in $M$, then any isometric embedding $i$ whose image lies in an $\mathbb{R}^3$ is admissible. Conditions (B) and (C)
 are necessary for the present proof of the positivity result Theorem 2 and they are general enough to include this most important case.
  We expect the most general positivity result holds under a convexity condition involving the Hamiltonian surface density 4-vector (2) of $\Sigma$ and the function $\tau$.

\section{Positivity of quasilocal energy}

We emphasize that although the definition of admissible pairs involves solving Jang's equation, our results only depend on the solvability and not on the specific solution. The expression of quasilocal energy only depends on the canonical gauge $\bar{u}^\nu$.

\textit{Theorem 2 - Suppose $M$ is a time-orientable spacetime that satisfies the dominant energy condition.
Suppose $\Sigma$ has spacelike mean curvature vector in $M$. Then the quasilocal energy (\ref{qlm_can}) with respect to any admissible pair $(i, t_0^\nu)$ is nonnegative.}

We take the time function $\tau$ on $i(\Sigma)$ and consider the Dirichlet problem of Jang's equation (\ref{jang1}) over $(\Omega, g_{ij}, K_{ij})$
 with $f=\tau$ on $\Sigma$. Condition (B) guarantees the equation is solvable on $\Omega$. Denote by $\widetilde{\Omega}$ the graph of the solution of Jang's equation. Schoen and Yau \cite{sy3} showed that if $M$ satisfies the dominant energy condition, there exists a vector filed $X$ on $\widetilde{\Omega}$ such that \begin{equation}\label{jang_scalar} R\geq 2|X|^2-2div X\end{equation} where $R$ is the scalar curvature of $\widetilde{\Omega}$.

 Let $\widetilde{\Sigma}$ be the graph of $\tau$ over $\Sigma$, and denote the outward normal of $\widetilde{\Sigma}$ with respect to
  $\widetilde{\Omega}$ by $\tilde{v}^i$ and the mean curvature of $\widetilde{\Sigma}$ with respect to $\tilde{v}^i$ by $\tilde{k}$.
  The boundary calculation in \cite{wy2} shows that $\sqrt{1+|\nabla\tau|^2}(\tilde{k}-\tilde{v}^iX_i)$ is equal to the expression in (\ref{pos_int}),
  and thus Condition (C) guarantees $\tilde{k}-\tilde{v}^iX_i>0$. We make use of another important property of the canonical gauge that
\begin{equation}\int_{\widetilde{\Sigma}} (\tilde{k}-\tilde{v}^i X_i)\geq
-\int_{\Sigma} [\bar{k}\bar{u}^\nu+ \bar{v}^\mu(\bar{K}^{\nu}_\mu-\bar{K}\delta^{\nu}_\mu)](t_0)_\nu.\end{equation} This is why the eventual
 definition of the quasilocal energy is independent of the solution of Jang's equation. On the other hand, it is not hard to check
 that $-\int_{\Sigma}[{k}_0{u_0}^\nu+ {v_0}^\mu({K_0}^{\nu}_\mu-{K_0}\delta^{\nu}_\mu)](t_0)_\nu =\int_{\widehat{\Sigma}} \widehat{k}$.
 (We were motivated by Gibbon's paper \cite{gi} to study this expression.) Here $\widehat{\Sigma}$ is the image of the projection of $i(\Sigma)$
 onto the orthogonal complement of $t_0^\nu$, and $\widehat{k}$ is the mean curvature of $\widehat{\Sigma}$. Therefore, the proof is reduced to the
  inequality $\int_{\widehat{\Sigma}} \widehat{k}\geq \int_{\widetilde{\Sigma}} (\tilde{k}-\tilde{v}^i X_i)$. We note that the Riemannian metrics
   on $\widetilde{\Sigma}$ and $\widehat{\Sigma}$ are the same. The proof will be  completed by the following comparison theorem in \cite{wy2} for the solution of Jang's equation.
%\begin{thm}\label{jang_sol}
Suppose $\widetilde{\Omega}$ is a Riemannian three-manifold with boundary $\widetilde{\Sigma}$, and suppose there exists a vector field $X$ on $\widetilde{\Omega}$ such that (\ref{jang_scalar}) holds on $\widetilde{\Omega}$
and $\tilde{k}>\tilde{v}^i X_i$ on $\widetilde{\Sigma}$. If the Gauss curvature of $\widetilde{\Sigma}$ is positive, and $k_0$ is the mean
curvature of the isometric embedding of $\widetilde{\Sigma}$ into $\mathbb{R}^3$, then
$\int_{\widetilde{\Sigma}} k_0  \geq \int_{\widetilde{\Sigma}} (\tilde{k}-\tilde{v}^i X_i)$.
%\end{thm}

\section{Definition of quasilocal mass and its positivity}

Our definition of quasilocal mass is similar to recovering the rest mass of a particle from the energy as measured by all observers of unit 4-velocities. The \textit{quasilocal mass} of $\Sigma$ in $M$ is defined to be the infimum of the quasilocal energy (\ref{qlm_can}) among all admissible pairs $(i, t_0^\nu)$.

Under the assumptions of Theorem 2, we obtain

\textit{Theorem 3 - If the set of admissible pairs is nonempty, then the quasilocal mass of $\Sigma$ in $M$ is nonnegative. In particular, this is the case when $\Sigma$ has positive Gauss curvature.}

The first part is clear from the definition. When $\Sigma$ has positive Gauss curvature, we can use Weyl's isometric embedding theorem to embed $\Sigma$ into a flat space-slice $\mathbb{R}^3$ on which the time function in $\mathbb{R}^{3,1}$ is a constant. Thus the admissible set is non-empty by Remark 1. This completes the proof of
Theorem 3.

Suppose the infimum is achieved by an admissible pair $(i, t_0^\nu)$, the \textit{quasilocal energy-momentum 4-vector} is defined as $m(\Sigma) t_0^\nu$ where $m(\Sigma)$ is the quasilocal mass of $\Sigma$. Therefore Theorem 3 implies the quasilocal energy-momentum 4-vector is always future-pointing and non-spacelike whenever it is defined.

\section{Properties of the new quasilocal mass}
  Expression (\ref{qlm_can}) contains the desired correction term; so the examples of surfaces in $\mathbb{R}^{3,1}$ found in \cite{ost} have zero
  quasilocal mass. The new quasilocal mass given in the previous section has the following properties:

\noindent 1. Suppose $\Sigma$ is a spacelike 2-surface which bounds a spacelike hypersurface $\Omega$ in a spacetime $M$.
The quasilocal mass is defined when the mean curvature vector of $\Sigma$ in $M$ is spacelike and the definition is independent of the choice of $\Omega$.
If $M$ satisfies the dominant energy condition and $\Sigma$ has positive intrinsic curvature, then the quasilocal mass is nonnegative. More generally, this holds if the set of admissible pairs for $\Sigma$ is nonempty.

\noindent 2. Any spacelike 2-surface in $\mathbb{R}^{3,1}$ with convex shadow in a time-direction (see Definition 1) has zero quasi-local mass.

\noindent 3. The small sphere limits of the quasilocal mass recover the matter energy-momentum tensor in the presence of matter and the Bel-Robinson tensor in vacuo, and the large sphere limits approach the ADM mass in the asymptotically flat case and the Bondi mass in the asymptotically null case.

We remark that the admissible pairs form an open subset of the set of functions $\tau$ on $\Sigma$ that satisfies (\ref{convex}).
The condition that the admissible set is nonempty in Theorem 3 is  a very mild assumption, and the quasilocal mass should be positive regardless of the sign of the intrinsic curvature of $\Sigma$.
The Euler-Lagrange equation for the energy minimizing isometric
    embedding into $\mathbb{R}^{3,1}$ is derived in \cite{wy2}.
When a $\Sigma$ in
  spacetime is given, we can solve for this equation and define the quasilocal energy-momentum 4-vector as in VI. The monotonicity property of our mass
  will be discussed in a forthcoming paper.


\end{document}